%% file: datta_ckm_proc.tex
\documentclass[12pt]{article}
\usepackage{epsfig}

\input econfmacros.tex
\def\Title#1{\begin{center} {\Large {\bf #1} } \end{center}}

\def\beq{\begin{equation*}}
\def\eeq{\end{equation*}}
\def\bea{\begin{eqnarray}}
\def\eea{\end{eqnarray}}
\def\roughly#1{\mathrel{\raise.3ex\hbox
{$#1$\kern-.75em\lower1ex\hbox{$\sim$}}}}

\def\bs{B_s^0}

\def\barp{{\raise.35ex\hbox
{${\sss (}$}}---{\raise.35ex\hbox{${\sss )}$}}}
\def\bdbarp{\hbox{$B_d$\kern-1.4em\raise1.4ex\hbox{\barp}}}
\def\bbarp{\hbox{$B$\kern-1.1em\raise1.4ex\hbox{\barp}}}
\def\sss{\scriptscriptstyle}

  \def\rr2{{1\over\sqrt{2}}}

\def\sss{\scriptscriptstyle}

\def\bs{B_s^0}

\def\roughly#1{\mathrel{\raise.3ex\hbox
{$#1$\kern-.75em\lower1ex\hbox{$\sim$}}}}

\def\btos{{\bar b} \to {\bar s}}

\def\bvv{B \to V_1 V_2}

\def\bsqq{b \to s \bar{q} q}

\def\BKstar{B\to\phi K^*}

\begin{document}
\begin{flushright}
UMISS-HEP-2012-09\\
Proceedings of CKM 2012, the 7th International Workshop on the CKM Unitarity \\Triangle,  University of Cincinnati, USA, 28 September - 2 October 2012.
\end{flushright}

\Title{$B \to VV$  Decays - Polarization and Triple Products.}

\bigskip\bigskip


\begin{raggedright}  

{\it Alakabha Datta\index{Datta, A.}\\
Department of Physics and Astronomy\\
University of Mississippi\\
University, MS, USA}
\bigskip\bigskip
\end{raggedright}

%
\begin{abstract}
\noindent We discuss how triple product asymmetries can be used to discover and constrain new physics in $\bvv$ decays.
\end{abstract}

\section{Introduction}

One of the main goals in $B$ physics experiments is to find new physics (NP) by
observing deviations from the standard model (SM) predictions \cite{dattaBnp}. 
The $\btos$ transitions are interesting as the SM CP violation  in these decays is tiny. 
Hence these decays are good places to search for  NP.
Two main probes of CP violation (CPV) are direct CP violation (DCPV) and indirect CP violation.
In $\bvv$ decays there is another probe of CPV- the triple product asymmetries (TPA) \cite{val, dattalondon, rosner}. As shown in \cite{dattalondon} they can provide useful information about the structure of NP.  
In this talk I will discuss TPA  in the the rare processes $\bsqq$, where $q$ are light quarks,  that lead to VV final states.

$\bvv$ decays are really three transitions because there are 3 polarization final states.
One can  construct direct CP violation (DCPV) asymmetries by taking the rate differences of the various polarization amplitudes. Interference between the different polarization amplitudes produce TPA. As is well known,
 DCPV $ \sim \sin{ \phi}\sin{ \delta}$ where $ \phi$ and $\delta$ are the weak and strong phase differences. TPA on the other hand are proportional to
$ \sim \sin{ \phi}\cos{ \delta}$. Hence DCPV and TPA complement each other. If the strong phases are small then TPA are maximized. 
There is another measurement involving triple products that is not CP violating which is called the fake triple product \cite{dattaTPnew}. This quantity goes as
$ \sim  \cos{ \phi}\sin{ \delta}$ and requires tagging for measurement.
This observable can constrain NP if the NP has the same weak phase as the SM in which  case DCPV and TPA vanish.

We will first discuss triple products in $\bvv$ decays in general. To motivate new physics we will consider the polarization measurements in $\btos$ transitions
that differ significantly from naive standard model (SM) predictions. There are SM solutions
to these polarization puzzles which can be tested \cite{dattapoltest} but we will focus on the NP solutions and  discuss how triple product asymmetries can constrain the size and structure of this new physics. Finally, we will discuss  polarization fractions and triple products in decays where the final states can be reached by both $B$  and $ \overline{B}$ decays and so mixing effects have to be included.

\section{ Triple Products}
In the $B$ rest frame we can construct a triple product , $TP$ 
\bea
T.P  &= &\vec{p} \cdot (\vec{\epsilon}_1\times
 \vec{\epsilon}_2), \
\eea
where $\vec{\epsilon}_{1,2}$ are the polarization vectors of the final state vector mesons and
 $\vec{p}$ is  the three momentum  of one of the vector mesons in the $B$ rest frame.
We can define a T-odd asymmetry
\bea
A_{T} & =  &\frac{\Gamma[T.P > 0]-\Gamma[T.P < 0]}{\Gamma[T.P > 0]+\Gamma[T.P < 0]}. \
\eea 
{}For true CP violation, we need to compare $A_{T}$ and $\bar{A}_{T}$. One can define the true and the fake TPA as 
\bea
A_{T.P}^{true} &= & A_{T}+\bar{A}_{T}\propto \sin{\phi} \cos{\delta},\nonumber\\
A_{T.P}^{fake} &= & A_{T}-\bar{A}_{T} \propto \cos{\phi} \sin{\delta}. \
\eea
 The TPA appear in the angular distribution of $\bvv \rightarrow(V_1 \rightarrow P_1P_1^\prime)
(V_2 \rightarrow P_2P_2^\prime)$. We can define two T.P's
\bea
A_T^{(1)} \equiv \frac{{\rm Im}(A_\perp
  A_0^*)}{A_0^2+A_\|^2+A_\perp^2} ~~,~~~~
A_T^{(2)} \equiv \frac{{\rm Im}(A_\perp
A_\|^*)}{A_0^2+A_\|^2+A_\perp^2} ~.
\eea 
Here the amplitudes are  longitudinal ($A_0$), or transverse to the
directions of motion and parallel ($A_\|$) or perpendicular
($A_\perp$) to one another.

For the CP conjugate decay one  defines two TPA
\bea
\bar{A}_T^{(1)} \equiv -\frac{{\rm Im}(\bar{A}_\perp
  \bar{A}_0^*)}{\bar{A}_0^2+\bar{A}_\|^2+\bar{A}_\perp^2} ~~,~~~~
\bar{A}_T^{(2)} \equiv - \frac{{\rm Im}(\bar{A}_\perp
\bar{A}_\|^*)}{\bar{A}_0^2+\bar{A}_\|^2+\bar{A}_\perp^2} ~.
\eea 

\section{ Testing NP with Triple Products}
In the SM there is prediction for one of the triple products in the heavy $b$ quark limit for
$B$ decays to light final states.
Let us write the transverse amplitudes in terms of the helicity amplitudes, $A_{\pm}$
\bea
A_\| &=& \frac{1}{\sqrt{2}} (A_+ + A_-) ~, \nonumber\\
A_\perp &=& \frac{1}{\sqrt{2}} (A_+ - A_-) ~.
\eea
Due to the fact that the weak
interactions are left-handed, the
helicity amplitudes obey the hierarchy 
\bea
\left\vert \frac{A_+}{A_-} \right\vert & = & r_T \sim  \frac{\Lambda_{QCD}}{m_b} ~.
\eea
Thus, in the heavy-quark limit, $A_\| = -A_\perp$ which means $A_T^{(2)}$, which is
proportional to ${\rm Im}(A_\perp A_\|^*)$, vanishes \cite{dattaTPnew} and
so  we can conclude that  the corresponding fake  and true TPA vanish  in
this limit.
 Corrections to the heavy quark limit can be calculated and within QCD factorization  this is at most around  10 \% \cite{dattaTPnew}. The other triple product $A_T^{(1)}$ does not vanish in the heavy quark limit and can be sizeable.

 In table~\ref{Kstphipoldata} the measurements for $B_d \to \phi
K^{*0}$ polarization observables \cite{hfag}  are shown.
\begin{table}
\centering
\begin{tabular}{|l|l|}\hline
\multicolumn{2}{|c|}{Polarization fractions}\\ \hline
 $f_L = 0.480 \pm 0.030$ &  $f_\perp = 0.241 \pm 0.029$  \\ \hline
\multicolumn{2}{|c|}{Phases}\\ \hline
$\phi_\parallel(rad) = 2.40^{+0.14}_{-0.13}$ & $\phi_\perp(rad) =2.39 \pm 0.13$ \\ \hline
$\Delta \phi_\parallel(rad) = 0.11 \pm 0.13$ & $\Delta \phi_\perp(rad) = 0.08 \pm 0.13$ \\ \hline
\multicolumn{2}{|c|}{CP asymmetries}\\ \hline
$A^0_{CP} = 0.04 \pm 0.06$ & $A^\perp_{CP} = -0.11 \pm 0.12$ \\ \hline
\end{tabular}
\caption{$B_d \to \phi K^{*0}$ polarization observables .}
\label{Kstphipoldata}
\end{table}
Using the numbers in the table we
can calculate the fake and true TPA:
\bea
A_{T.P, 2}^{fake} &=& \frac{1}{2}(A_{T,B}^{(2)} - \bar{A}_{T,{\bar{B}}}^{(2)} )= 0.002\pm 0.049 ~,\nonumber\\
A_{T.P, 1}^{fake} &=&\frac{1}{2}(A_{T,B}^{(1)} - \bar{A}_{T,{\bar{B}}}^{(1)} )= -0.23 \pm 0.03 ~.\
\eea
The measured values of the fake TPA are  in agreement with the
SM prediction in the heavy quark limit.  
 The true T.P are
\bea
A_{T.P, 2}^{true} &=& \frac{1}{2}(A_{T,B}^{(2)} + \bar{A}_{T,{\bar{B}}}^{(2)} )= -0.004\pm 0.025,\nonumber\\ 
A_{T.P, 1}^{true} &=& \frac{1}{2}(A_{T,B}^{(1)} + \bar{A}_{T,{\bar{B}}}^{(1)} )= 0.013\pm 0.053.\
\eea
These are consistent with SM or with NP with the same weak phase as the SM amplitude.

If one assumes NP is responsible for the large transverse polarization ($f_T$) observed in penguin/penguin dominated decays like $\BKstar$ then the NP operators must have the structures \cite{dattapol} 
 \bea
S_{LL}=(1-\gamma_5) \otimes (1-\gamma_5) \quad or \quad T_{LL} = \sigma^{ \mu \nu} (1-\gamma_5) \otimes \sigma_{\mu \nu} (1-\gamma_5),\nonumber\\
S_{RR}=(1+\gamma_5) \otimes (1+\gamma_5) \quad or \quad  T_{RR} = \sigma^{\mu \nu} (1+\gamma_5) \otimes \sigma_{\mu \nu} (1+\gamma_5).\
\eea
With the assumption  $f_T^{SM}=0$ one can draw the following conclusions.
 In the heavy-quark limit, $A_+ = 0$ for the $LL$ operators, so that
$A_\| = -A_\perp$ (as in the SM) and $A_T^{(2)} = 0$ which in turn implies the corresponding  TPA vanish.
 Similarly, for the
$RR$ operators  $A_- = 0$, so that $A_\| = A_\perp$,
$A_T^{(2)} = 0$ and the corresponding  TPA vanish. 
However both $LL$ and $RR$ operators cannot be present.
If the SM produces a large $f_T$ from penguin annihilation and/or re scattering( which is left handed) then the
$RR$ operator cannot be present.
Thus, the measurement of $A_T^{(2)} \simeq
0$ rules out ${RR}$ operators, or at least strongly constrains
them.

\section{Time Dependent Angular Distribution}
We consider decays like $B_s \to \phi\phi( \bar{b} \to \bar{s} s \bar{s}),
 K^*\bar{K}^*( \bar{b} \to \bar{s} d \bar{d}) $. Here
 the final state can be reached by both $B_{s}$ and $\overline{B}_{s}$ decays so mixing effects have to be included \cite{dattaVVnew}.
 Assuming that
$V_{1,2}$ both decay into pseudoscalars (i.e.\ $V_1 \to P_1 P_1'$,
$V_2 \to P_2 P_2'$), the angular distribution of the decay is 
given in terms of the vector $\vec{\omega} = (\cos{\theta_1
},\cos{\theta_2 }, \Phi)$ 
\bea
\frac{d^3 \Gamma(t)}{d\vec{\omega}} &=& \frac{9}{32 \pi} \sum^6_{i=1} K_i(t) f_i(\vec{\omega}) ~.
\eea
Here, $\theta_1 (\theta_2 )$ is the angle between the directions of motion of  $P_1 (P_2 )$ in the $V_1 (V_2 ) $ rest frame and  $V_1 (V_2 ) $ in the $B$ rest frame, and $\Phi$ is the 
angle between
the normals to the planes defined by $P_1 P_1^{\prime}$ and $P_2 P_2^{\prime}$ in the $B$ rest frame. 
The functions $K_i(t)$ are expressed in terms of the $B_s$ oscillation parameters, $\phi_s$ ,  $\Gamma_s$ , 
$\Delta \Gamma_s$, $\Delta m_s$ and the  transversity amplitudes $A_{i (= 0, \parallel, \perp)}$ \cite{dattaVVnew}.  The
time-integrated untagged angular distribution can be obtained by
integrating the $K_i(t)+\bar{K}_i(t)$ observables over time:
\bea
\label{ADphiphi-inte}
\frac{d^3 \langle\Gamma(\bs \to f) \rangle}{ d\vec{\omega} } &=&
\frac{9}{32 \pi} \sum^6_{i=1} \langle K_i \rangle f_i(\vec{\omega})~,
\eea
where
\bea
\label{TIKis}
\langle\Gamma(\bs \to f) \rangle   &=& \frac{1}{2}\int^\infty_0 dt (\Gamma^{B_s} + \Gamma^{\bar{B}_s})~,\quad
\langle K_i \rangle = \frac{1}{2}\int^\infty_0 dt(K_i(t)+\bar{K}_i(t))~.
\eea
The general structure is
$$\langle K_i \rangle \propto {\cal{A}}^{ch}_i + {\cal{A}}^{sh}_i y_s$$
where $y_s={\Delta \Gamma_s \over 2 \Gamma_s}$.
The ${\cal{A}}^{ch}_i$ can be used to extract the polarization fractions and triple products.
The details can be found in Ref.~\cite{dattaVVnew}. The key point is that the TPA can be measured with untagged time integrated measurements.

\section{Acknowledgements}  This work was financially supported  in part by the National Science Foundation under Grant No.\ NSF PHY-1068052.

\end{document}

%% file: econfmacros.tex
\def\beq{\begin{equation}}
\def\eeq#1{\label{#1}\end{equation}}
\def\eeqn{\end{equation}}

\def\beqa{\begin{eqnarray}}
\def\eeqa#1{\label{#1}\end{eqnarray}}
\def\eeqan{\end{eqnarray}}

\let\bar=\overbar

\def\Dslash{\not{\hbox{\kern-4pt $D$}}}
\def\dslash{\not{\hbox{\kern-2pt $\del$}}}

\def\msb{{\bar{\ssstyle M \kern -1pt S}}}

